\documentclass[5p,sort&compress]{elsarticle}
\usepackage{graphicx,lineno,hyperref}
\usepackage[version=3]{mhchem}
\usepackage{graphicx,amssymb,amsmath,textcomp}
\usepackage[T1]{fontenc}
\usepackage{color}
\usepackage[dvipsnames]{xcolor}
\usepackage{chemfig}

\journal{Colloid and Interfacial Science Communications}

\begin{document}

\begin{frontmatter}

\title{Intentionally added ionic surfactants induce\\ Jones-Ray effect at air-water interface}
\author[kyushu]{Yuki Uematsu}
\ead{uematsu@chem.kyushu-univ.jp}
\author[kyushu]{Kengo Chida}
\author[kyushu]{Hiroki Matsubara}
\address[kyushu]{Department of Chemistry, Kyushu University, Fukuoka 819-0395, Japan}
\date{\today}

\begin{abstract}
The Jones-Ray effect is an anomalous minimum in the surface tension of aqueous electrolytes at millimolar salt concentrations. 
We experimentally demonstrated that intentionally added ionic surfactants induce the Jones-Ray effect. 
The one-dimensional Poisson-Boltzmann theory, including the effect of surfactant adsorption and salt depletion, excellently agrees with the obtained experimental data.
All the parameters of the ion-specific surface affinities used in the theory are consistent with previous experiments.
These results strongly suggest that the Jones-Ray effect observed so far has been induced by the inevitable contamination of the air-water interfaces.
\end{abstract}


\end{frontmatter}

\newpage
One of the central issues in the physical chemistry of interfaces is the surface charge of the air-water interface \cite{Schelero2011,Douwe2017}.
The surface tension of air-electrolyte interfaces is helpful to understand the surface affinity of ions \cite{Washburn1930,OnsagerSamaras1934,MatubayasiBook}.
Most inorganic electrolytes increase the surface tension, reflecting that the ions are surface-inactive \cite{OnsagerSamaras1934}.
However, the surface tension of the electrolytes shows a characteristic minimum at millimolar electrolyte concentration values when the concentration is varied, known as the Jones-Ray effect \cite{JonesRay1937,JonesRay1941-2}.
The Jones-Ray effect has been controversial, but it has been experimentally reproduced by different methods and by different research groups \cite{DoleSwartout1940,Passoth1959,Randles1966,Pivat2006,Pivat2010,Roke2016,Sangwai2017,Sangwai2018}.
The Jones-Ray effect was first explained by anion adsorption \cite{Dole1937}, which was supported by surface-sensitive nonlinear spectroscopy \cite{Petersen2004}.
However, although large halides, such as iodide, are absorbed into the air-water interface, the Jones-Ray effect is independent of the type of the anion present.
In fact, NaCl causes the Jones-Ray effect even though the chloride ion is known to be surface-inactive \cite{JonesRay1941-2}.
Additionally, Langmuir has explained the Jones-Ray effect by the reduction of the capillary radius because of the wetting films in the capillary rise method \cite{Langmuir1938-1}.
However, this mechanism does not work for the du No\"uy ring \cite{DoleSwartout1940}, maximum bubble pressure \cite{Passoth1959,Randles1966}, and Wilhelmy plate methods \cite{Pivat2006,Pivat2010,Roke2016}, which were also used to observe the Jones-Ray effect.
Another explanation for the Jones-Ray effect is $\mathrm{OH^-}$ adsorption \cite{KarrakerRadke2002,ManciuRuckenstein2003}.
This scenario is supported by the negative zeta potential of bubbles in water \cite{Smith1985} and the repulsive disjoining pressure of wetting films on silica \cite{Schelero2011}.
However, this scenario contradicts with the surface tension of NaOH solutions because the surface tension of NaOH solutions is larger than that of NaCl solutions, meaning that $\mathrm{OH^-}$ is more surface-inactive than $\mathrm{Cl^-}$ \cite{Washburn1930,Uematsu2018,Douwe2017}. 
Although other explanations were suggested \cite{NicholsPratt1984,Roke2017}, the experimental fact that the concentration and depth of the minima depend on independent measurements \cite{JonesRay1937,JonesRay1941-2,DoleSwartout1940,Passoth1959,Randles1966,Pivat2006,Pivat2010,Roke2016} implies uncontrolled hidden parameters in the experiments.
Recently, one of the authors constructed the theory of the surface tension of electrolytes, and the Jones-Ray effect is explained by the traces of charged impurities in the water \cite{Uematsu2018,Uematsu2018-2}. 
In this scenario, assuming that the charged impurities have an affinity similar to typical ionic surfactants, namely sodium dodecylsulfate, the sufficient concentration of the impurities is a few nanomolar, which cannot be detected by conductivity measurements.
This implies that the minima observed by Jones and Ray were caused by inevitable and unexpected contamination of the solutions.

Jones and Ray, using the capillary rise method, measured the surface tension of the solutions with an accuracy of approximately $\pm0.001\,$mN/m. 
However, this accuracy level requires a special apparatus, and such a tremendous precision has been difficult to achieve so far.
According to the impurity scenario \cite{Uematsu2018}, intentionally added ionic surfactants make the surface tension minimum deeper and the concentration of the minimum larger.  
Although the surface tension of electrolytes in the presence of ionic surfactants has been studied to date, these studies have not focused on the Jones-Ray effect \cite{TanakaIkeda1991,Warszyski1998,Kralchevsky1999}.
Fixing ionic surfactants at millimolar concentrations, the surface tension is a decreasing function of the salt concentration \cite{TanakaIkeda1991,Warszyski1998,Kralchevsky1999}.
Because the surface tension of inorganic salt solutions without ionic surfactants is normally an increasing function of the salt concentration, there must be a crossover from an increasing function to a decreasing function.
At the crossover, anomalous behavior such as minima and maxima could be observed \cite{TanakaIkeda1991}.
Therefore, in this letter, the surface tensions of salt solutions in the presence of ionic surfactants at micromolar concentrations are measured by the drop-volume method, which is easy to handle with an accuracy of approximately $\pm0.05\,$mN/m.
In this communication, combining experiment and theory, we demonstrate that intentionally added ionic surfactants at micromolar concentrations induce a minimum in the surface tension at approximately $10\,$mM($=10\,$mmol/dm$^3$).
These results strongly support the previous theory claiming that the Jones-Ray effect is caused by the contamination of the solutions with charged impurities.

\begin{figure}
\center
\includegraphics[width=0.5\textwidth]{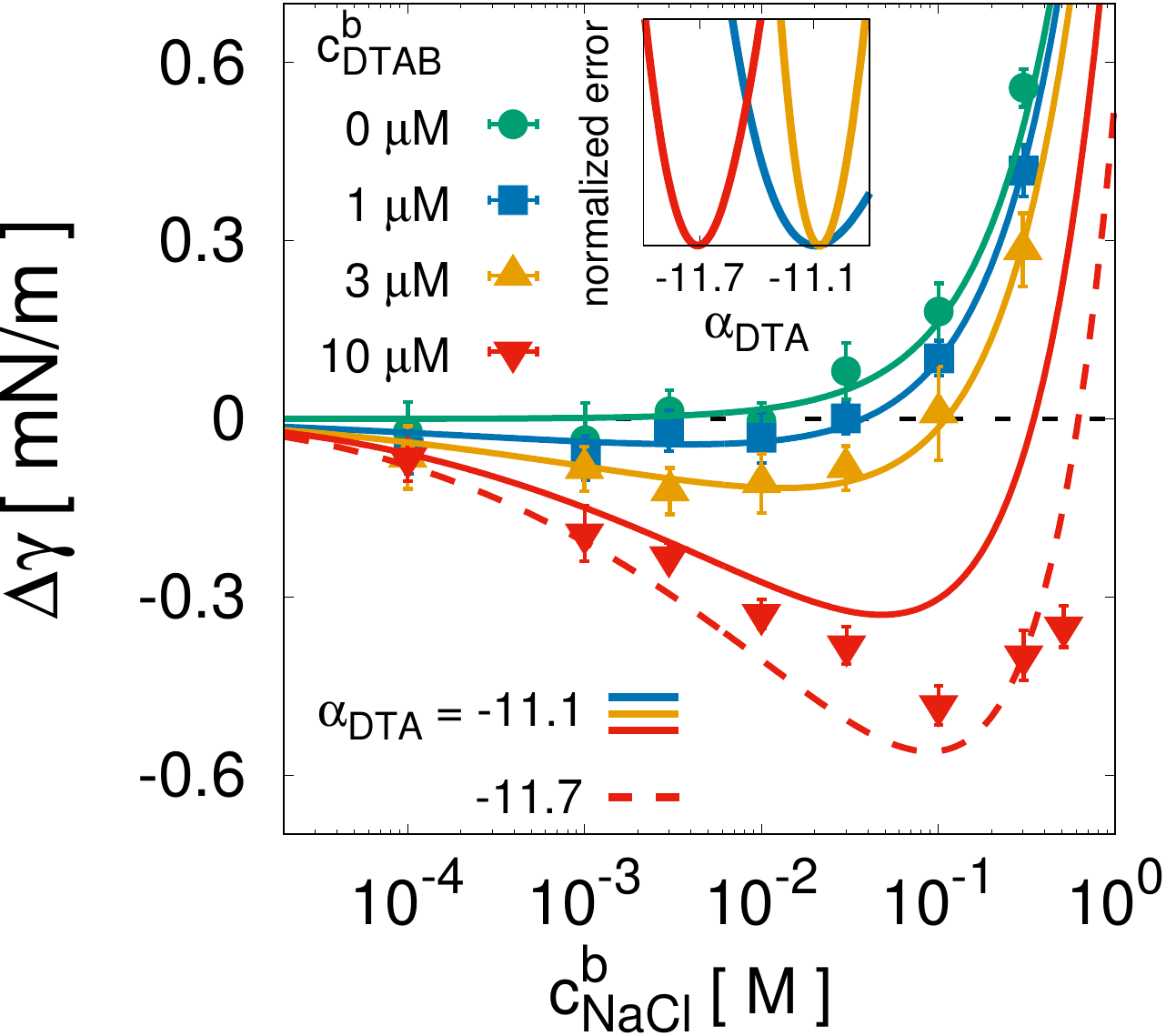}
\caption{
Surface tension of the NaCl solutions in the absence and presence of DTAB.
The points are the experimental data and the lines are the theoretical calculations.
For the theoretical calculations, we used $T=298\,$K, $\varepsilon=78$, $z^*=0.5\,$nm, $\alpha_\mathrm{Na}=1.2$, $\alpha_\mathrm{Cl}=1.0$, and $\alpha_\mathrm{Br}=0.4$.
For the blue, yellow, and red solid lines, we used $\alpha_\mathrm{DTA}=-11.1$, and for the red broken line, we used $\alpha_\mathrm{DTA}=-11.7$.
The inset shows the errors of the least-squares fits normalized by their minimum with varying $\alpha_\mathrm{DTA}$, for the DTAB concentration $c^\mathrm{b}_\mathrm{DTAB}=1\,\mu$M (blue), $3\,\mu$M (yellow), and $10\,\mu$M (red).}
\label{fig:1}
\end{figure}

\begin{figure*}
\center
\includegraphics[width=0.8\textwidth]{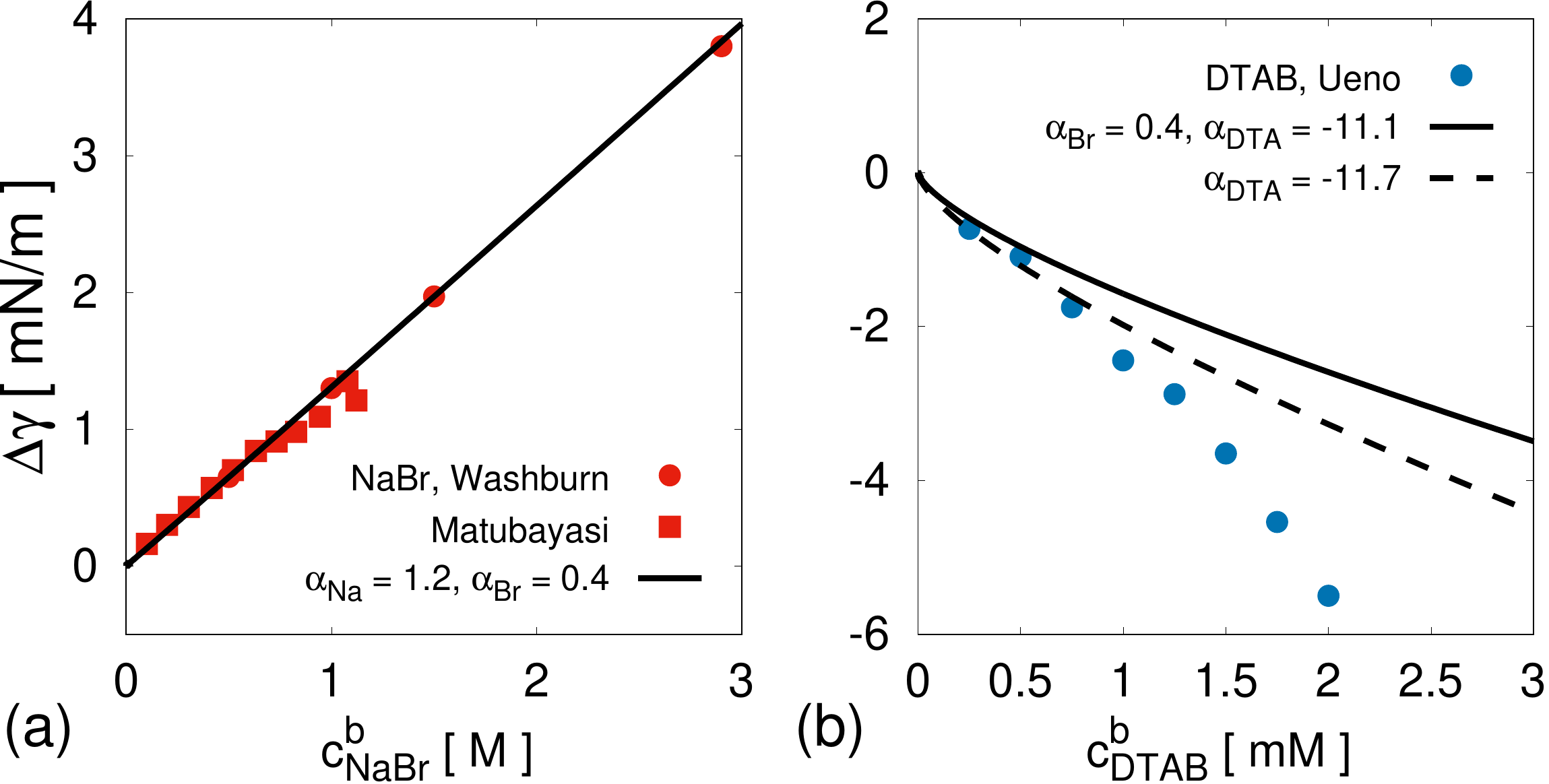}
\caption{
(a) The surface tension of the NaBr solutions.
The points are extracted from the literatures \cite{Washburn1930,MatubayasiBook} whereas the lines are theoretically calculated by adjusting $\alpha_\mathrm{Br}=0.4$ and using $\varepsilon=78$, $T=298\,$K, $z^*=0.5\,$nm, and $\alpha_\mathrm{Na}=1.2$.
(b) The surface tension of the DTAB solutions.
The points are extracted from the literature \cite{Ueno2018}, whereas the solid and broken lines are theoretically calculated by using $\alpha_\mathrm{DTA}=-11.1$ and $-11.7$, and with fixing $\alpha_\mathrm{Br}=0.4$.
Parts of the experimental data in (a) and (b) are provided in units of mol/kg, but we neglect the small difference between the units of mol/kg and M.}
\label{fig:2}
\end{figure*}

Figure~\ref{fig:1} shows the excess surface tension of the NaCl solutions in the absence and presence of dodecyltrimethylammonium bromide (DTAB), a cationic surfactant.
The green points are the data of the NaCl solution without DTAB.
Although a small negative surface tension $\Delta\gamma=-0.04\pm0.06\,$mN/m is observed at $c^\mathrm{b}_\mathrm{NaCl}=1\,$mM, we think that this is an experimental error and not the Jones-Ray effect because the minimum observed by Jones and Ray at $1\,$mM is approximately $\Delta\gamma\approx-0.01\,$mN/m \cite{JonesRay1937,JonesRay1941-2}. 
The blue, yellow, and red points denote the surface tension data as a function of the NaCl concentration $c^\mathrm{b}_\mathrm{NaCl}$ in the presence of DTAB at the concentrations of $c^\mathrm{b}_\mathrm{DTAB}=1\,\mu$M, $3\,\mu$M, and $10\,\mu$M, respectively.
We find that when the concentration of DTAB increases, the minimum appears at approximately $c^\mathrm{b}_\mathrm{NaCl}=10\,$mM.
For the DTAB concentration at $c^\mathrm{b}_\mathrm{DTAB}=10\,\mu$M, the minimum is distinct and located at approximately $c^\mathrm{b}_\mathrm{NaCl}=100\,$mM.

By using the mean-field theory for the surface tension of the air-electrolyte interface \cite{Uematsu2018}, we quantitatively explain the experimental data.
This theory is based on the Poisson-Boltzmann equation with the ion-specific adsorption potentials such as
\begin{equation}
\varepsilon\varepsilon_0\frac{d^2\psi}{dz^2}=-e\sum_i q_ic_i^\mathrm{b}\mathrm{e}^{-q_ie\psi/k_\mathrm{B}T-\alpha_i\theta(z^*-z)},
\label{eq:1}
\end{equation}
where $\varepsilon$ is the dielectric constant, $\varepsilon_0$ is the vacuum permittivity, $\psi$ is the local electrostatic potential, $z$ is the coordinate normal to the surface, $e$ is the elementary charge, $q_i$ is the valency of the $i$ type ion, $c_i^\mathrm{b}$ is the bulk concentration of the $i$ type ion, $k_\mathrm{B}T$ is the thermal energy, $\alpha_i$ is the adsorption energy of the $i$ type ion, $\theta(z)$ is the Heaviside step function, and $z^*$ is the thickness of the interfacial adsorption potentials.
This adsorption energy $\alpha_i$ is positive for most inorganic ions and negative for ionic surfactants \cite{Uematsu2018}.
Here we consider four types of ions: $\mathrm{Na^+}$, $\mathrm{Cl^-}$, dodecyltrimethylammonium ion ($\mathrm{DTA^+}$), and $\mathrm{Br^-}$, and we neglect $\mathrm{H_3O^+}$, $\mathrm{OH^-}$, $\mathrm{HCO_3^-}$, and $\mathrm{CO_3^{2-}}$ because their concentrations are dilute.
The surface tension is obtained by integration of the Gibbs isotherm equation given by
\begin{equation}
d\gamma = -\sum_i \Gamma_i d\mu_i,
\label{eq:2}
\end{equation}
where $\Gamma_i=c^\mathrm{b}_i\int^\infty_0(\mathrm{e}^{-eq_i\psi(z)/k_\mathrm{B}T-\alpha_i\theta(z^*-z)}-1)dz$ is the surface excess of the type $i$ ion, and $\mu_i$ is the chemical potential of the type $i$ ion. 
In the same way as Ref.~\citenum{Uematsu2018}, we use the ideal-gas approximation, $d\mu_i=k_\mathrm{B}Tdc^\mathrm{b}_i/c^\mathrm{b}_i$, $T=298\,$K, $\varepsilon=78$, and $z^*=0.5\,$nm.  
For Na$^+$ and Cl$^-$, we use $\alpha_\mathrm{Na}=1.2$ and $\alpha_\mathrm{Cl}=1.0$, respectively, which are extracted from the potentials of mean force calculated by molecular dynamics simulations \cite{Horinek2009}.
The green line in Figure~\ref{fig:1} is the surface tension of the NaCl solution as a function of $c^\mathrm{b}_\mathrm{NaCl}(=c^\mathrm{b}_\mathrm{Na}=c^\mathrm{b}_\mathrm{Cl}$) with $\alpha_\mathrm{Na}=1.2$ and $\alpha_\mathrm{Cl}=1.0$, in agreement with the experimental data (green points).
To determine the affinity of Br$^-$, we fit $\alpha_\mathrm{Br}$ by using the surface tension data of NaBr \cite{Washburn1930, MatubayasiBook}.
When we use $\alpha_\mathrm{Br}=0.4$ with fixing $\alpha_\mathrm{Na}=1.2$, a good agreement is obtained, as shown in Figure~\ref{fig:2}a.

Next, we fit the theory with the obtained experimental data by adjusting the surface affinity of $\mathrm{DTA^+}$, $\alpha_\mathrm{DTA}$. 
We optimized $\alpha_\mathrm{DTA}$ by the least-squares method with each data set, where the normalized errors are plotted in the inset of Figure~\ref{fig:1}.
Both the data sets $c^\mathrm{b}_\mathrm{DTAB}(=c^\mathrm{b}_\mathrm{DTA}=c^\mathrm{b}_\mathrm{Br})=1\,\mu$M and $3\,\mu$M give the same surface affinity of $\alpha_\mathrm{DTA}=-11.1$.
However, for $c^\mathrm{b}_\mathrm{DTAB}=10\,\mu$M, the calculation with $\alpha_\mathrm{DTAB}=-11.1$ (red solid line) underestimates the minimum depth, and we obtain a different optimized surface affinity of $\alpha_\mathrm{DTA}=-11.7$ (red broken lines).
This means that the surfactant adsorption energy is no longer constant for such a large DTAB concentration. 

Furthermore, we examine the obtained DTAB surface affinities, $\alpha_\mathrm{DTA}=-11.1$ and $-11.7$ to compare with the surface tension data of the DTAB solution \cite{Ueno2018}.
In Figure~\ref{fig:2}b, the solid line is the calculation with $\alpha_\mathrm{DTA}=-11.1$ and $\alpha_\mathrm{Br}=0.4$, which is in fair agreement with the first two experimental data values. 
The broken line is the calculation with $\alpha_\mathrm{DTA}=-11.7$, which agrees with one point more than the solid line.
Such a small difference between $\alpha_\mathrm{DTA}=-11.1$ and $-11.7$ is reasonable because the theory uses the ideal-gas approximation. 
For more precise analysis, we need to include the attractive interactions between surfactants expressed by the Frumkin adsorption isotherm or others \cite{Kralchevsky1999}. 

\begin{figure}
\center
\includegraphics[width=0.5\textwidth]{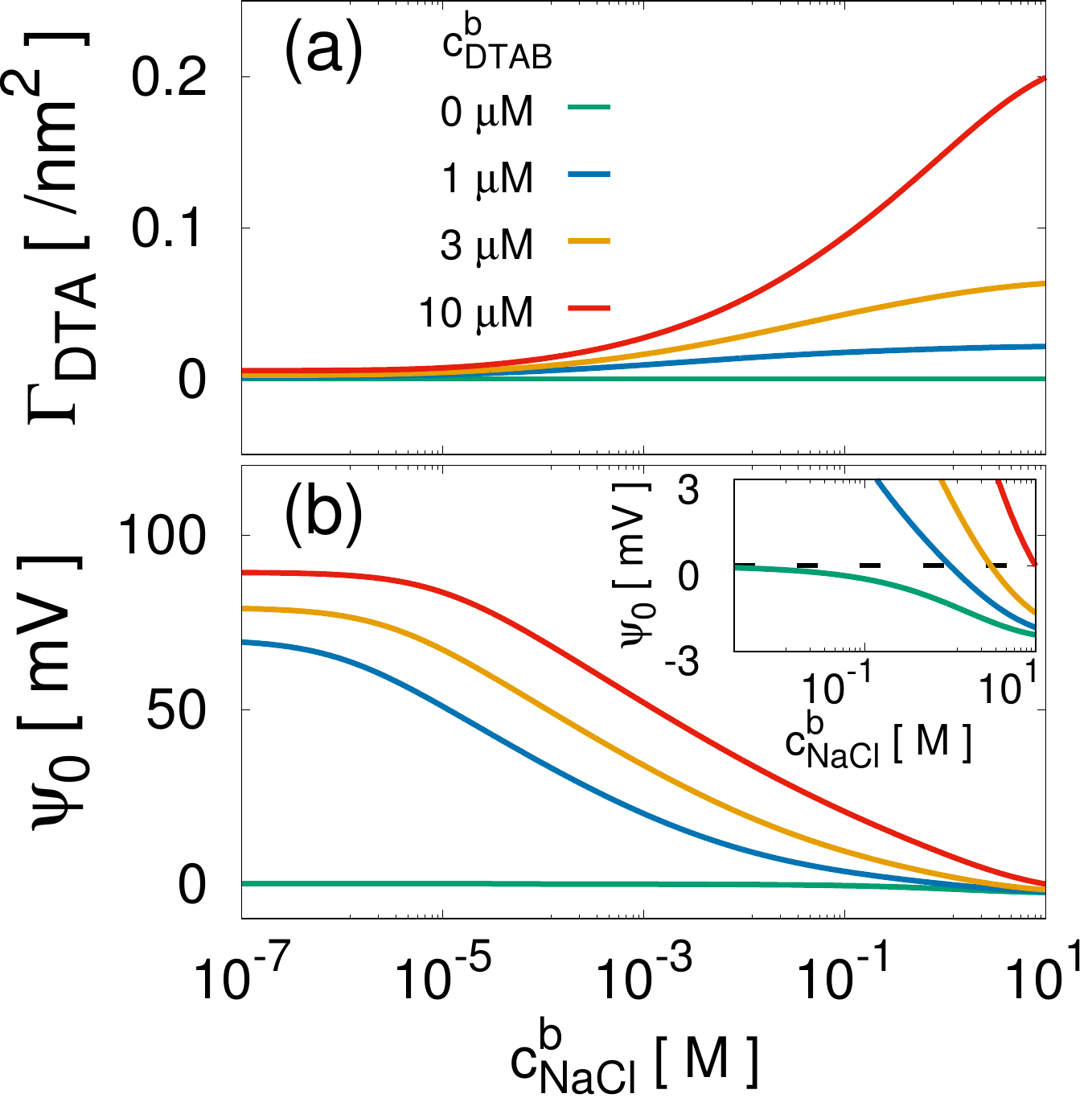}
\caption{
(a) Surface excess of DTA$^+$, $\Gamma_\mathrm{DTA}$, as a function of the NaCl concentration.
For the theoretical calculation we used $T=298\,$K, $\varepsilon=78$, $z^*=0.5\,$nm, $\alpha_\mathrm{Na}=1.2$, $\alpha_\mathrm{Cl}=1.0$, $\alpha_\mathrm{Br}=0.4$, and $\alpha_\mathrm{DTA}=-11.1$.
(b) Surface potential $\psi_0$ as a function of the NaCl concentration. 
We used the same parameters as in (a). 
The inset is an enlarged plot showing small negative surface potentials for high $c^\mathrm{b}_\mathrm{NaCl}$.
}
\label{fig:3}
\end{figure}

From the theoretical analysis, the surface excess of $\mathrm{DTA^+}$ and the surface potential $\psi_0=\psi|_{z=0}$ can be obtained as shown in Figure~\ref{fig:3}. 
As shown in Figure~\ref{fig:3}a, the surface excess, $\Gamma_\mathrm{DTA}$ is an increasing function as $c^\mathrm{b}_\mathrm{NaCl}$ increases.
The typical value for high $c^\mathrm{b}_\mathrm{NaCl}$ is $0.1\,$/nm$^2$ which equals one $\mathrm{DTA^+}$ molecule per $10\,$nm$^2$.
This increase of $\Gamma_\mathrm{DTA}$ originates from the salt-induced screening of the electrostatic repulsion between the ionic surfactants. 
The adsorption of the ionic surfactants is compensated with the adsorption of $\mathrm{Cl^-}$, as well as $\mathrm{Br^-}$, because of the charge neutrality $\sum_iq_i\Gamma_i=0$.
Thus, considering the Gibbs adsorption isotherm, eq.~(\ref{eq:2}), this is the reason why the minimum appears in the surface tension.
As shown in Figure~\ref{fig:3}b, the surface potential $\psi_0$ is positive and approximately $100\,$mV for low $c^\mathrm{b}_\mathrm{NaCl}$. 
The sign of the surface potential is determined by the sign of the ionic surfactants. 
This large potential is suppressed by the salt, and for large $c^\mathrm{b}_\mathrm{NaCl}$, a small negative potential appears (the inset of Figure~\ref{fig:3}b) because the depletion of $\mathrm{Na^+}$ is slightly stronger than the depletion of $\mathrm{Cl^-}$ ($\alpha_\mathrm{Na}<\alpha_\mathrm{Cl}$).
Note that these positive surface potentials are observed in the surface potential measurements of surfactant solutions \cite{Warszyski1998} and the zeta potential measurements of air bubbles in cationic surfactant solutions \cite{Smith1985}. 

In summary, we measure the surface tension of the NaCl solutions in the absence and presence of DTAB at micromolar concentrations.
Without DTAB, the surface tension curves, as a function of the NaCl concentration, do not show an apparent minimum.
This does not conflict with the Jones-Ray effect because the precision of the measurements is not sufficient to detect a small minimum, approximately $\Delta\gamma\approx -0.01\,$mN/m.
In the presence of DTAB at micromolar concentrations, we obtained a minimum large enough to be observed by the drop-volume method.
This minimum can be quantitatively explained by the Poisson-Boltzmann theory with the ion-specific adsorption potentials. 
We fitted only the adsorption energy of $\mathrm{DTA^+}$ and obtained excellent agreement between experiment and theory. 
The resultant adsorption energy of $\mathrm{DTA^+}$ also quantitatively agrees with the surface tension data of the DTAB solutions at very low concentrations, showing that all the parameters in the theory are consistent.
The mechanism of the surface tension minimum is the adsorption of the ionic surfactants because of the screening of the electrostatic repulsion between ionic surfactants at the surface.
These results strongly support that the Jones-Ray effect is caused by traces of charged impurities in water \cite{Uematsu2018}.
Contamination of the air-water interface plays an important role also in other interfacial anomalies, such as the disjoining pressure of wetting films \cite{Schelero2011}, zeta potential of hydrophobic surfaces \cite{Smith1985}, and hydrodynamic boundary conditions of air-water interfaces \cite{Manor2008-2}.
We hope that this communication will stimulate further physico-chemical research in these fields.

{\it Experimental methods}.---NaCl was of high grade, 99.99 \% (Merck Millipore), and was used without further purification. 
DTAB was of grade 99\%(Wako Pure Chemical Industries, Japan) and was purified by recrystallizing it five times from an acetone/ethanol mixture (volume ratio 5/1).
Water was Elix pure water (conductivity $>5.0\,$M$\Omega$cm) produced by Elix Advantage 5 (Merck Millipore).
The temperature was set at $25.0\,$\textcelsius.
The solutions at various concentrations were prepared by dilution.

\begin{figure}
\center
\includegraphics[width=0.45\textwidth]{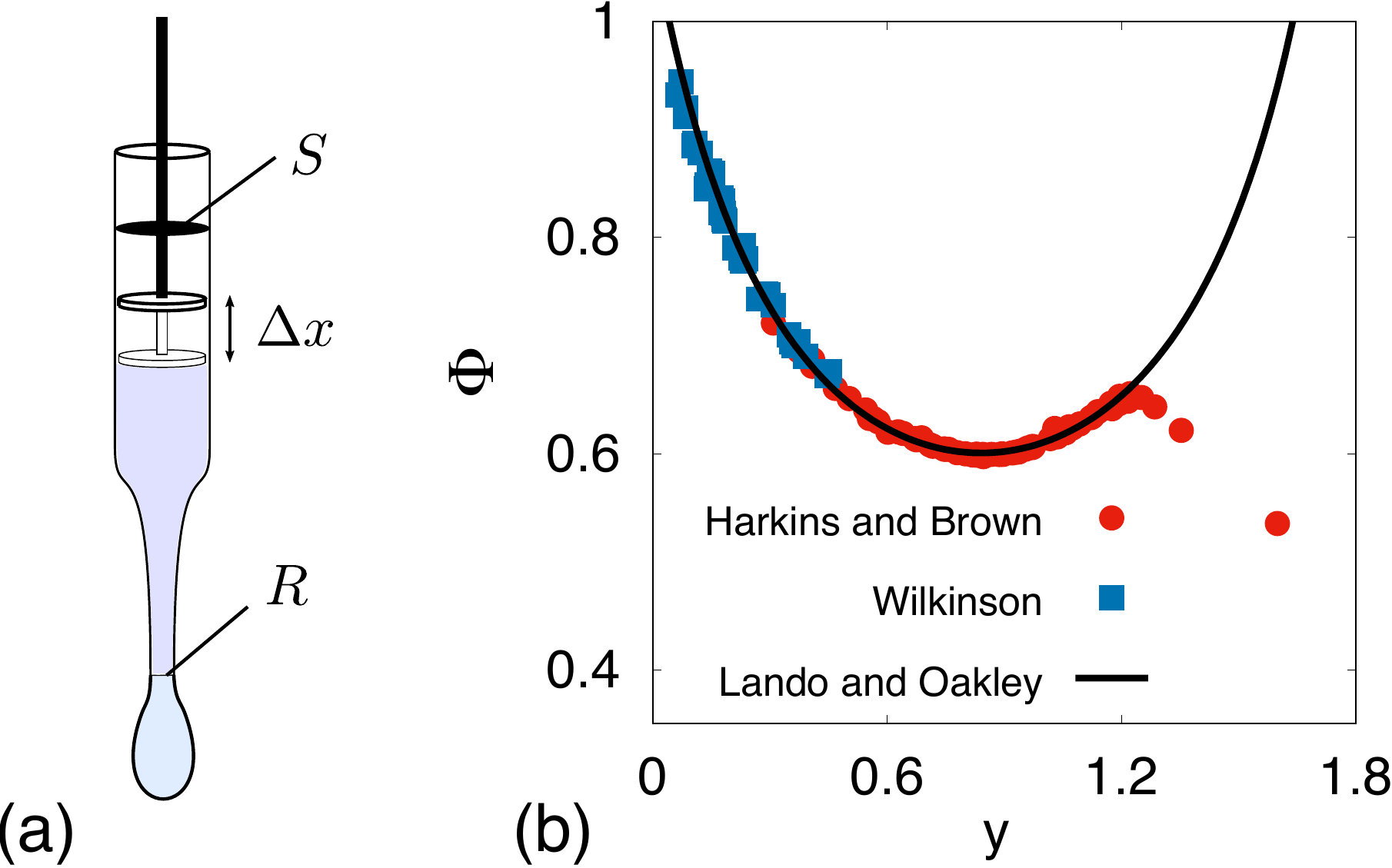}
\caption{
(a) Schematic illustration of the drop-volume method.
$S$ is the cross section of the syringe, $\Delta x$ is the moving distance of the plunger, and $R$ is the outer radius of the tip.
(b) The function $\Phi(y)$ as a function of $y = R/(S\Delta x)^{1/3}$.
The red points \cite{Harkins1919} and the blue points \cite{Wilkinson1972} are the experimental data, whereas the solid line is eq.~(\ref{eq:4}) \cite{Lando1967}. 
}
\label{fig:4}
\end{figure}

The surface tension was measured by the drop-volume method (DSV2000, Yamashita Giken, Japan).
The drop-volume method is a technique to measure the surface tension of gas-liquid or liquid-liquid interfaces.
Measuring the volume of a drop, which just falls from the horizontal tip of the apparatus because of the gravity, gives the surface tension as shown in Figure~\ref{fig:4}a. 
The surface tension $\gamma$ is given by \cite{Harkins1919}
\begin{equation}
\gamma=\frac{S\,\Delta x\, \Delta\rho\, g}{2\pi R\,\Phi(y)},
\label{eq:3}
\end{equation}
where $S$ is the cross section of the syringe, $\Delta x$ is the moving distance of the plunger, $\Delta\rho = \rho_\mathrm{sol}-\rho_\mathrm{air}$ is the density difference between solution and air, $g$ is the acceleration of the gravity, $R$ is the outer radius of the tip, and $\Phi(y)$ is an empirical function of $y=R/(S\Delta x)^{1/3}$ (see Figure~\ref{fig:4}b).
We used an approximated polynomial function for $\Phi(y)$ \cite{Lando1967}
\begin{equation}
\frac{1}{2\pi\Phi(y)}=0.14782 +0.27896y-0.166y^2,
\label{eq:4}
\end{equation}
which is the solid line in Figure~\ref{fig:4}b that is in agreement with the experimental data (blue and red points) \cite{Harkins1919,Wilkinson1972} for $0.05\lesssim y\lesssim 1.2$.
We used the standard gravity $g=9.80665\,$m/s$^2$, the cross section $S=12.566\,$mm$^2$, the normal air density at $25\,$\textcelsius$\,$ and $1\,$atm $\rho_\mathrm{air}=0.001184\,$g/mL, and a practical formula for the NaCl solution density \cite{JonesRay1941-2},
\begin{equation} 
\rho_\mathrm{sol}=0.997074 + 0.041882c^\mathrm{b}_\mathrm{NaCl} - 0.001878{c^\mathrm{b}_\mathrm{NaCl}}^{3/2},
\label{eq:5}
\end{equation}
where $c_\mathrm{NaCl}^\mathrm{b}$ is the molar concentration in unit M and $\rho_\mathrm{sol}$ is in unit g/mL.
We used eq.~(\ref{eq:5}) for solutions containing even a small amount of DTAB.
The outer radius of the tip $R$ is determined as $R=1.1515\,$mm from the surface tension measurement of pure water by setting $\gamma=71.96\,$mN/m \cite{MatubayasiBook} in eq.~(\ref{eq:3}).
In the experiment, we waited for $2\,$min after forming a droplet likely to fall at the tip, and then, the plunger was pushed at $2\,\mu$m/sec until the droplet fell.
We measured the moving distance of the plunger $\Delta x$ at least $12\,$times for each data point. 
Because we obtained $\Delta x\approx 3\,$mm and $R\approx 1\,$mm for all measurements, the resultant $y$ is $y=R/(S\Delta x)^{1/3}\approx0.3$, which is applicable to eq.~(\ref{eq:4}).

\section*{Acknowledgements}
We acknowledge Xiaolei Xu for giving the purified DTAB crystals.
Y.U.~was supported by Grant-in-Aid for JSPS Fellows 16J00042.

\section*{References}
\bibliography{tension.bib}

\begin{thebibliography}{10}
\expandafter\ifx\csname url\endcsname\relax
  \def\url#1{\texttt{#1}}\fi
\expandafter\ifx\csname urlprefix\endcsname\relax\def\urlprefix{URL }\fi
\expandafter\ifx\csname href\endcsname\relax
  \def\href#1#2{#2} \def\path#1{#1}\fi

\bibitem{Schelero2011}
N.~Schelero, R.~von Klitzing, Correlation between specific ion adsorption at
  the air/water interface and long-range interactions in colloidal systems,
  Soft Matter 7 (2011) 2936.
\newblock \href {http://dx.doi.org/10.1039/C0SM01162B}
  {\path{doi:10.1039/C0SM01162B}}.

\bibitem{Douwe2017}
S.~Mamatkulov, C.~Allolio, R.~R. Netz, D.~J. Bonthuis, Orientation induces
  adsorption of the hydrated proton at the air-water interface, Angew. Chem.
  Int. Ed. 56 (2017) 15846--15851.
\newblock \href {http://dx.doi.org/10.1002/anie.201707391}
  {\path{doi:10.1002/anie.201707391}}.

\bibitem{Washburn1930}
E.~W. Washburn (Ed.), International Critical Tables of Numerical Data, Physics,
  Chemistry and Technology, electronic 1st Edition, Vol.~IV, Knovel, Norwich,
  NY, 1930.

\bibitem{OnsagerSamaras1934}
L.~Onsager, N.~N.~T. Samaras, The surface tension of {D}ebye-{H}\"uckel
  electrolytes, J. Chem. Phys. 2 (1934) 528--536.
\newblock \href {http://dx.doi.org/10.1063/1.1749522}
  {\path{doi:10.1063/1.1749522}}.

\bibitem{MatubayasiBook}
N.~Matubayasi, Surface Tension and Related Thermodynamic Quantities of Aqueous
  Electrolyte Solutions, CRC Press, 2014.

\bibitem{JonesRay1937}
G.~Jones, W.~A. Ray, The surface tension of solutions of electrolytes as a
  function of the concentration. {I}. {A} differential method for measuring
  relative surface tension, J. Am. Chem. Soc. 59 (1937) 187--198.
\newblock \href {http://dx.doi.org/10.1021/ja01280a048}
  {\path{doi:10.1021/ja01280a048}}.

\bibitem{JonesRay1941-2}
G.~Jones, W.~A. Ray, The surface tension of solutions of electrolytes as a
  function of the concentration {III}. {S}odium chloride, J. Am. Chem. Soc. 63
  (1941) 3262--3263.
\newblock \href {http://dx.doi.org/10.1021/ja01857a007}
  {\path{doi:10.1021/ja01857a007}}.

\bibitem{DoleSwartout1940}
M.~Dole, J.~A. Swartout, A twin-ring surface tensiometer. {I}. {T}he apparent
  surface tension of potassium chloride solutions, J. Am. Chem. Soc. 62 (1940)
  3039--3045.
\newblock \href {http://dx.doi.org/10.1021/ja01868a042}
  {\path{doi:10.1021/ja01868a042}}.

\bibitem{Passoth1959}
G.~Passoth, {\"U}ber den {J}ones-{R}ay-{E}ffekt und die {O}berfl\"achenspannung
  verd\"unnter {E}lektrolytl\"osungen, Z. Physik. Chem. 211O (1959) 129--147.
\newblock \href {http://dx.doi.org/10.1515/zpch-1959-21116}
  {\path{doi:10.1515/zpch-1959-21116}}.

\bibitem{Randles1966}
J.~E.~B. Randles, D.~J. Schiffrin, Surface tension of dilute acid solutions,
  Trans. Faraday Soc. 62 (1966) 2403--2408.
\newblock \href {http://dx.doi.org/10.1039/TF9666202403}
  {\path{doi:10.1039/TF9666202403}}.

\bibitem{Pivat2006}
A.~Acharid, M.~Sadiki, G.~Elmanfe, N.~Derkaoui, R.~Olier, M.~Privat, Water
  behavior revisited at the air/ionic solution and silica/ionic solution
  interfaces. extension to coadsorption of ions and organic molecules, Langmuir
  22 (2006) 8790--8799.
\newblock \href {http://dx.doi.org/10.1021/la061160c}
  {\path{doi:10.1021/la061160c}}.

\bibitem{Pivat2010}
A.~Azri, G.~Elmanfe, R.~Olier, M.~Privat, Environmental copper: Behaviour when
  involved in physical adsorption at several interfaces, Atmospheric
  Environment 44 (2010) 5211--5217.
\newblock \href {http://dx.doi.org/10.1016/j.atmosenv.2010.08.054}
  {\path{doi:10.1016/j.atmosenv.2010.08.054}}.

\bibitem{Roke2016}
Y.~Chen, H.~I. Okur, N.~Gomopoulos, C.~Macias-Romero, P.~S. Cremer, P.~B.
  Petersen, G.~Tocci, D.~M. Wilkins, C.~Liang, M.~Ceriotti, S.~Roke,
  Electrolytes induce long-range orientational order and free energy changes in
  the {H}-bond network of bulk water, Sci. Adv. 2 (2016) e1501891.
\newblock \href {http://dx.doi.org/10.1126/sciadv.1501891}
  {\path{doi:10.1126/sciadv.1501891}}.

\bibitem{Sangwai2017}
A.~Kakati, J.~S. Sangwai, Effect of monovalent and divalent salts on the
  interfacial tension of pure hydrocarbon-brine systems relevant for low
  salinity water flooding, J. Petro. Sci. Eng 157 (2017) 1106--1114.
\newblock \href {http://dx.doi.org/10.1016/j.petrol.2017.08.017}
  {\path{doi:10.1016/j.petrol.2017.08.017}}.

\bibitem{Sangwai2018}
N.~K. Jha, S.~Iglauer, J.~S. Sangwai, Effect of monovalent and divalent salts
  on the interfacial tension of n-heptane against aqueous anionic surfactant
  solutions, J. Chem. Eng. Data 63 (2018) 2341--2350.
\newblock \href {http://dx.doi.org/10.1021/acs.jced.7b00640}
  {\path{doi:10.1021/acs.jced.7b00640}}.

\bibitem{Dole1937}
M.~Dole, Surface tension of strong electrolytes, Nature 140 (1937) 464--465.
\newblock \href {http://dx.doi.org/10.1038/140464b0}
  {\path{doi:10.1038/140464b0}}.

\bibitem{Petersen2004}
P.~B. Petersen, J.~C. Johnson, K.~P. Knutsen, R.~J. Saykally, Direct
  experimental validation of the {J}ones-{R}ay effect, Chem. Phys. Lett. 397
  (2004) 46--50.
\newblock \href {http://dx.doi.org/10.1016/j.cplett.2004.08.048}
  {\path{doi:10.1016/j.cplett.2004.08.048}}.

\bibitem{Langmuir1938-1}
I.~Langmuir, Repulsive forces between charged surfaces in water and the cause
  of the {J}ones-{R}ay effect, Science 88 (1938) 430--432.
\newblock \href {http://dx.doi.org/10.1126/science.88.2288.430}
  {\path{doi:10.1126/science.88.2288.430}}.

\bibitem{KarrakerRadke2002}
K.~A. Karraker, C.~J. Radke, Disjoining pressures, zeta potentials and surface
  tensions of aqueous non-ionic surfactant electrolyte solutions: Theory and
  comparison to experiment, Adv. Colloid Int. Sci. 96 (2002) 231--264.
\newblock \href {http://dx.doi.org/10.1016/S0001-8686(01)00083-5}
  {\path{doi:10.1016/S0001-8686(01)00083-5}}.

\bibitem{ManciuRuckenstein2003}
M.~Manciu, E.~Ruckenstein, Specific ion effects via ion hydration: {I}.
  {S}urface tension, Adv. Colloid Int. Sci. 105 (2003) 63--101.
\newblock \href {http://dx.doi.org/10.1016/S0001-8686(03)00018-6}
  {\path{doi:10.1016/S0001-8686(03)00018-6}}.

\bibitem{Smith1985}
N.~P. Brandon, G.~H. Keksall, S.~Levine, A.~L. Smith, Interfacial electrical
  properties of electrogenerated bubbles, J. Appl. Electrochem. 15 (1985)
  485--493.
\newblock \href {http://dx.doi.org/10.1007/BF01059289}
  {\path{doi:10.1007/BF01059289}}.

\bibitem{Uematsu2018}
Y.~Uematsu, D.~J. Bonthuis, R.~R. Netz, Charged surface-active impurities at
  nanomolar concentration induce {J}ones-{R}ay effect, J. Phys. Chem. Lett. 9
  (2018) 189--193.
\newblock \href {http://dx.doi.org/10.1021/acs.jpclett.7b02960}
  {\path{doi:10.1021/acs.jpclett.7b02960}}.

\bibitem{NicholsPratt1984}
A.~L. Nichols, L.~R. Pratt, Salt effects on the surface tensions of dilute
  electrolyte solutions: The influence of nonzero relative solubility of the
  salt between the coexisting phases, J. Chem. Phys. 80 (1984) 6225.
\newblock \href {http://dx.doi.org/10.1063/1.446725}
  {\path{doi:10.1063/1.446725}}.

\bibitem{Roke2017}
H.~I. Okur, Y.~Chen, D.~M. Wilkins, S.~Roke, The {J}ones-{R}ay effect
  reinterpreted: Surface tension minima of low ionic strength electrolyte
  solutions are caused by electric field induced water-water correlations,
  Chem. Phys. Lett. 684 (2017) 433--442.
\newblock \href {http://dx.doi.org/10.1016/j.cplett.2017.06.018}
  {\path{doi:10.1016/j.cplett.2017.06.018}}.

\bibitem{Uematsu2018-2}
Y.~Uematsu, D.~J. Bonthuis, R.~R. Netz, Impurity effects at hydrophobic
  surfaces, Current Opinion in Electrochemistry XX (2018) xx--xx.
\newblock \href {http://dx.doi.org/10.1016/j.coelec.2018.09.003}
  {\path{doi:10.1016/j.coelec.2018.09.003}}.

\bibitem{TanakaIkeda1991}
A.~Tanaka, S.~Ikeda, Adsorption of dodecyltrimethylammonium bromide on aqueous
  surfaces of sodium bromide solutions, Colloids and Surfaces 56 (1991)
  217--229.
\newblock \href {http://dx.doi.org/10.1016/0166-6622(91)80122-5}
  {\path{doi:10.1016/0166-6622(91)80122-5}}.

\bibitem{Warszyski1998}
P.~Warszy\'nski, W.~Barzyk, K.~Lunkenheimer, H.~Fruhner, Surface tension and
  surface potential of {N}a n-dodecyl sulfate at the air-solution interface:
  Model and experiment, J. Phys. Chem. B 102 (1998) 10948--10957.
\newblock \href {http://dx.doi.org/10.1021/jp983901r}
  {\path{doi:10.1021/jp983901r}}.

\bibitem{Kralchevsky1999}
P.~A. Kralchevsky, K.~D. Danov, G.~Broze, A.~Mehreteab, Thermodynamics of ionic
  surfactant adsorption with account for the counterion binding: Effect of
  salts of various valency, Langmuir 15 (1999) 2351--2365.
\newblock \href {http://dx.doi.org/10.1021/la981127t}
  {\path{doi:10.1021/la981127t}}.

\bibitem{Ueno2018}
S.~Ueno, Y.~Takajo, S.~Ikeda, R.~Takemoto, Y.~Imai, T.~Takiue, H.~Matsubara,
  M.~Aratono, Surface dilational viscoelasticity of aqueous surfactant
  solutions by surface quasi-elastic light scattering, Colloid and Polymer
  Sicence 296 (2018) 781--791.
\newblock \href {http://dx.doi.org/10.1007/s00396-018-4297-8}
  {\path{doi:10.1007/s00396-018-4297-8}}.

\bibitem{Horinek2009}
D.~Horinek, A.~Herz, L.~Vrbka, F.~Sedlmeier, S.~I. Mamatkulov, R.~R. Netz,
  Specific ion adsorption at the air/water interface: The role of hydrophobic
  solvation, Chem. Phys. Lett. 479 (2009) 173--183.
\newblock \href {http://dx.doi.org/10.1016/j.cplett.2009.07.077}
  {\path{doi:10.1016/j.cplett.2009.07.077}}.

\bibitem{Manor2008-2}
O.~Manor, I.~U. Vakarelski, G.~W. Stevens, F.~Grieser, R.~R. Dagastine,
  D.~Y.~C. Chan, Dynamic forces between bubbles and surfaces and hydrodynamic
  boundary conditions, Langmuir 24 (2008) 11533--11543.
\newblock \href {http://dx.doi.org/10.1021/la802206q}
  {\path{doi:10.1021/la802206q}}.

\bibitem{Harkins1919}
W.~D. Harkins, F.~E. Brown, The determination of surface tension (free surface
  energy), and the weight of falling drops: The surface tension of water and
  benzen by the capillary height method, J. Am. Chem. Soc. 41 (1919) 499--524.
\newblock \href {http://dx.doi.org/10.1021/ja01461a003}
  {\path{doi:10.1021/ja01461a003}}.

\bibitem{Lando1967}
J.~L. Lando, H.~T. Oakley, Tabulated correction factors for the
  drop-weight-volume determination of surface and interfacial tensions, J.
  Colloid Interface Sci. 25 (1967) 526--530.
\newblock \href {http://dx.doi.org/10.1016/0021-9797(67)90064-1}
  {\path{doi:10.1016/0021-9797(67)90064-1}}.

\bibitem{Wilkinson1972}
M.~C. Wilkinson, Extended use of, and comments on, the drop-weight
  (drop-volume) technique for the determination of surface and interfacial
  tensions, J. Colloid Interface Sci. 40 (1972) 14--26.
\newblock \href {http://dx.doi.org/10.1016/0021-9797(72)90169-5}
  {\path{doi:10.1016/0021-9797(72)90169-5}}.

\end{thebibliography}

\end{document}